# Magnon straintronics in the 2D van der Waals ferromagnet CrSBr from first-principles

Dorye L. Esteras, [‡,†] Andrey Rybakov, [‡,†] Alberto M. Ruiz,[†] José J. Baldoví[*,†]

[†]Instituto de Ciencia Molecular, Universitat de València, Catedrático José Beltrán 2, 46980 Paterna, Spain. E-mail: j.jaime.baldovi@uv.es

ABSTRACT: The recent isolation of two-dimensional (2D) magnets offers tantalizing opportunities for spintronics and magnonics at the limit of miniaturization. One of the key advantages of atomically-thin materials is their outstanding deformation capacity, which provides an exciting avenue to control their properties by strain engineering. Herein, we investigate the magnetic properties, magnon dispersion and spin dynamics of the air-stable 2D magnetic semiconductor CrSBr ($T_C$ = 146 K) under mechanical strain using first-principles calculations. Our results provide a deep microscopic analysis of the competing interactions that stabilize the long-range ferromagnetic order in the monolayer. We showcase that the magnon dynamics of CrSBr can be modified selectively along the two main crystallographic directions as a function of applied strain, probing the potential of this quasi-1D electronic system for magnon straintronics applications. Moreover, we predict a strain-driven enhancement of $T_C$ considering environmental screening by ~30%, allowing the propagation of spin waves at higher temperatures.

Magnonics is an emerging research field within nanomagnetism and nanoscience that investigates the transmission, storage, and processing of information using spin waves (SWs) as an alternative to conventional electronics.[1–3] The use of SWs, whose quanta are referred to as magnons, instead of transport of electric charges offers unique and compelling opportunities such as extremely low power consumption, shorter wavelengths, nanoscale devices, tunable spectrum and wave-based computing concepts to name a few.[4–7] Besides unlocking new horizons in fundamental physics, the recent discovery of long-range magnetic order in atomically-thin crystals provides an unprecedented platform for magnonics at the limit of miniaturization.[8,9] Among the family of 2D magnetic crystals, layered antiferromagnets such as $CrI_3$ or CrSBr are particularly interesting because their magnetic properties can be controlled by electrostatic doping,[10] electric fields[11] or strain,[12] can hold long-lived magnons in the GHz to THz range[13,14] and their van der Waals (vdW) nature ensures an easy transfer onto the surface of other nanomaterials in order to improve the device performance.[15]

Strain engineering has been shown to be a powerful tool to tune the lattice, electronic and magnetic properties of 2D materials due to their larger elasticity compared to bulk materials.[16–22] In intrinsic 2D magnets, strain has been able to induce switching between antiferromagnetic (AFM) and ferromagnetic (FM) interlayer coupling and successfully modulated magnetic exchange interactions, magnetic anisotropy and Curie temperature ($T_C$) down to the single-layer.[23–25] In the context of magnonics, a new branch called magnon straintronics has recently been proposed and experimentally implemented, showing the potential of strain tuning and SW coupling via magnon-phonon resonance to modulate dynamically SW transport, while offering a promising route to generate SWs.[26–28] However, the effect of strain on the SW dynamics of 2D materials is still unexplored –even from a theoretical point of view– and deserves urgent attention, owing to its potential to develop a new generation of 2D magnonic devices.

Motivated by these burgeoning developments, herein, we focus on the 2D air-stable semiconductor CrSBr, which is formed by ferromagnetic layers ($T_C \sim 150$ K) with antiferromagnetic interlayer coupling.[29,30] Interestingly, the material presents a quasi-1D electronic structure that is entangled with its magnetic structure, resulting in an ideal candidate to control the magnetic properties by applying uniaxial deformations of the lattice.[31] With the aim of investigating the effects of tensile and compressive strain on the SW dynamics of CrSBr at the 2D limit, we have implemented an efficient first-principles methodology that combines density functional theory (DFT), a derived tight-binding Hamiltonian, spin wave theory and atomistic simulations. This allows us to provide a detailed microscopic analysis of magnetic exchange interactions and thus the propagation of SWs. Interestingly, we show that magnons in single-layer CrSBr can be modified selectively along the two main crystallographic directions as a function of the applied strain, which paves the way to the use of this 2D semiconductor for magnon straintronics applications.

Bulk CrSBr crystallizes in the orthorhombic Pmmm space group with lattice parameters $a = 3.50$ Å, $b = 4.76$ Å and $c = 7.96$ Å, exhibiting a vdW layered structure.[32] This allows the crystal to be exfoliated down to single-layer flakes.[30] In each layer, the Cr atoms are embedded in a distorted octahedral coordination environment and are connected to their nearest-neighbor Cr atoms by sulfur and bromine atoms along the *a* axis, whereas only by sulfur atoms along the *b* and *c* axes. In contrast with CrI$_3$, the network of Cr magnetic atoms is not co-planar, spins are oriented in-plane along the *b* axis and the different length of the lattice vectors provides a marked structural anisotropy that results in decoupled quasi-1D chains along *a* and *b* as proved by conductivity measurements.[33] These features make this material unique in terms of structural properties among the 2D magnetic family.

In CrSBr monolayer, magnetic exchange interactions between Cr atoms can be modeled considering superexchange mechanisms through the *p* orbitals of Br and S ligands up to third nearest neighbors.[34] Thus, there are mainly three magnetic exchange interactions represented by $J_1$, $J_2$, $J_3$, where $J_1$ accounts for the interaction between Cr-Br-Cr (89º) and Cr-S-Cr (95º) atoms along *a* direction; $J_2$ defines coupling between Cr atoms from different "sublayers" along *c* (97º); and $J_3$ between Cr atoms mediated by softer S bridges along the *b* axis (160º) (Fig. 1). The intricate relation between magnetism and crystal lattice has been recently investigated by synchrotron X-ray diffraction measurements, evidencing that the thermal evolution of the lattice parameters follows intriguing opposing trends when cooling down the crystal.[31] While *a* tends to elongate due to an unconventional expansion when decreasing temperature from 260 K the so-called spin-freezing temperature T* ~40 K, both *b* and *c* lattice parameters become shorter. Intriguingly, these trends lead to a progressive enhancement of $J_1$, $J_2$ and $J_3$.[35]

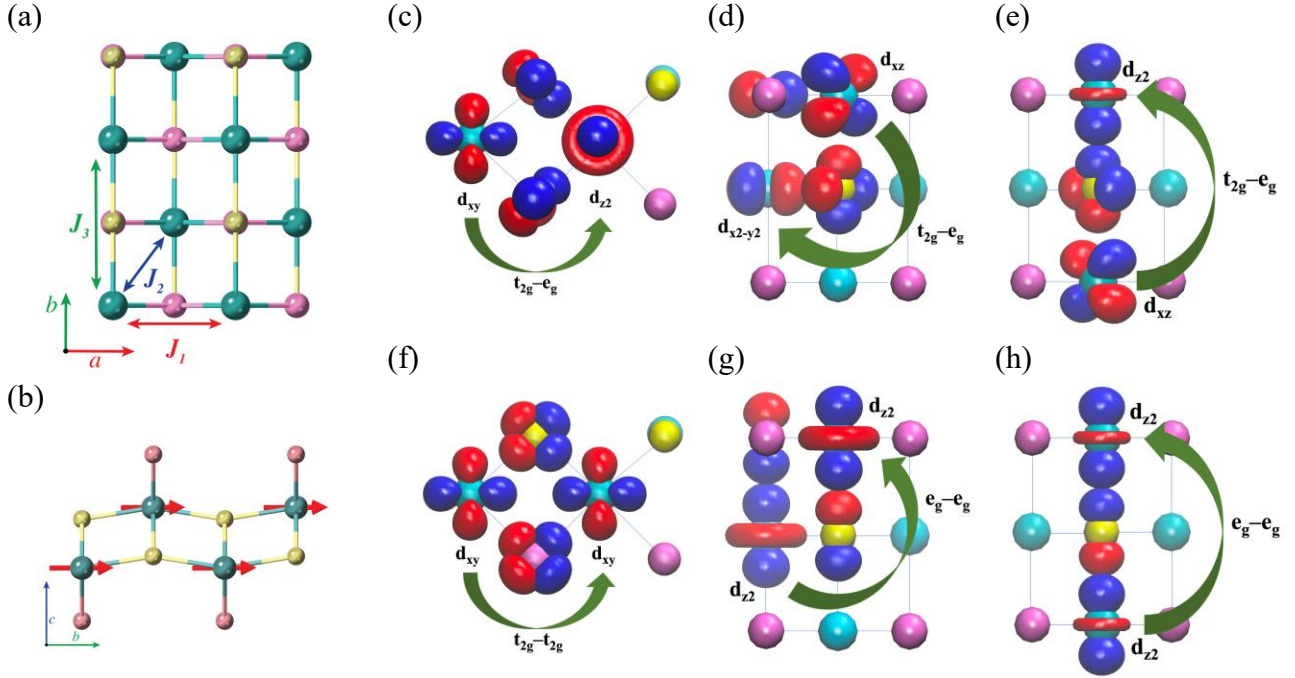

**Figure 1.** (a) Top view of the crystal structure of a single CrSBr layer. Cyan, yellow and pink balls represent chromium, sulfur and bromine atoms, respectively. $J_1$, $J_2$ and $J_3$ magnetic exchange interactions for first-, second- and third-neighbors are represented by arrows that connect the Cr atoms. (b) Side view of the same CrSBr structure showing the spin orientation along *b*. (c-h) Calculated maximally-localized Wannier orbitals. Green arrows illustrate the most relevant magnetic superexchange channels, namely $t_{2g}$-$e_g$ (FM), $t_{2g}$-$t_{2g}$ (AFM) and $e_g$-$e_g$ (AFM) for $J_1$ (c, f), $J_2$ (d, g) and $J_3$ (e, h).

In order to understand the electronic and magnetic structure of single-layer CrSBr, we perform first-principles calculations by means of Hubbard-corrected spin-polarized DFT including spin-orbit coupling (SOC), applying a Hubbard U = 3 eV to describe the highly localized *d* electrons of Cr (See methods). The DFT+U+SOC electronic band structure and density of states (DOS) are plotted in Fig. S1. One can observe the presence of highly dispersive conduction bands along the X-S and Y-Γ directions, which are characteristic of this 2D semiconductor with quasi-1D electronic properties.[33] The orbital resolved DOS (Fig. S1b) illustrates the key contribution of the *d* orbitals of Cr and *p* orbitals of S and Br atoms around the Fermi level. The octahedral crystal field around the Cr atoms splits the *d* orbitals into two sets of energy levels, namely $t_{2g}$ ($d_{xy}$, $d_{xz}$ and $d_{yz}$) and $e_g$ ($d_{z2}$ and $d_{x2-y2}$), leading to a conduction channel formed between spin up states.[36] In our calculations each Cr atom carries a net magnetic moment of 3.03 $\mu_B$, which agrees well with S= 3/2 for $Cr^{3+}$, while S and Br atoms are slightly spin polarized with -0.27 $\mu_B$ and -0.08 $\mu_B$, respectively. The most stable magnetic configuration is achieved for ferromagnetic intralayer ordering with in-plane easy axis of magnetization along the *b* axis (Fig. 1b), in agreement with experiments.[37–39]

To determine magnetic exchange interactions, we derive a tight-binding Hamiltonian expressed in the basis of maximally-localized Wannier functions (MLWFs)[40] including SOC that permits to employ the Green's function approach as implemented in the TB2J package.[41] This method treats the local spin rotation as a perturbation (See Supplementary Section, Methods) and is an efficient route that circumvents the main limitations of total energy mapping analysis, i.e. (i) the requirement of a number of magnetic configurations at least equal to the parameters of the Hamiltonian, (ii) the difficult convergence for some metastable configurations or (iii) the use of large supercells. The obtained magnetic exchange parameters for U=3 eV ($J_1$ = 3.54 meV, $J_2$ = 3.08 meV and $J_3$ = 4.15 meV) are close to the ones recently estimated from neutron diffraction measurements in bulk CrSBr,[42] and are in agreement with theoretical calculations for the monolayer reported in the literature (see Table S1).[43,44] We also provide the evolution of these parameters as a function of Hubbard U accompanied by the orbital decomposed magnetic exchange channels (Tables S2-S5; Fig. S2), which are graphically represented in Fig. 1 (c-h). One can observe that almost in the entire range of U, the three magnetic exchange parameters are > 0 due to the dominant $t_{2g}$-$e_g$ (FM) pathway; however, while $J_1$ increases with U, $J_2$ and $J_3$ decrease. This different evolution comes from the particular orbitals involved in the FM and AFM superexchange mechanisms for each direction. In the case of $J_1$, the AFM $t_{2g}$-$t_{2g}$ hoping takes place between the $d_{xy}$ of the two Cr atoms and $p_x$, $p_y$ orbitals of the ligands contained in the *ac* plane. It can be observed that an enhancement of the Coulomb interactions drastically limits the $t_{2g}$-$t_{2g}$ AFM pathway, which is relevant along *a*. By contrast, the dominant AFM exchange for $J_2$ and $J_3$ arises from $e_g$-like orbitals, mainly $d_{z^2}$ that points along *b*, whose occupations increase with U. This favors the AFM $d_{z^2}$-$p_z$-$d_{z^2}$ superexchange pathway along the *b* direction, whereas the $t_{2g}$-$e_g$ FM mechanism is slightly improved, leading to $J_3 < 0$ at U ~ 6 eV.

Then, we investigate the strain-dependent evolution of magnetic exchange parameters by simulating the application of two types of uniaxial strain (along perpendicular *a* and *b* axes). The lattice parameters were varied up to a 5% of compression and elongation. In parallel, Hubbard U was scanned in a range of 1-6 eV to rationalize the effect of different screening scenarios. To obtain a dense grid of outputs for the two variables –strain and Hubbard U–, we applied the least squares method using eq. 1, which successfully reproduces the calculated results and can be used to interpolate the intermediate points:

$$J = \sum_{i=0}^{3} \sum_{j=0}^{3} a_{ij} U^i \varepsilon^j \qquad (1)$$

where *J* are the magnetic exchange interactions up to three nearest neighbors, $a_{i,j}$ are the fitting coefficients, U is the on-site Hubbard parameter, $\varepsilon$ is % strain and i, j, are the powers of the fit.

Fig. 2 shows a 3D surface plot of the dependence of each J with respect to $\varepsilon$ and U. Due to the asymmetric structure of CrSBr, one can observe that each exchange interaction presents an independent evolution depending on the crystallographic direction along which strain is applied. As described above, $J_1$ and $J_3$ connect Cr atoms along the *a* and *b* axes, respectively. This causes that $J_1$ parameter is strongly influenced by uniaxial strain in *a*, whereas $J_3$ changes notably by applying strain along *b*. On the other hand, J$_2$ interaction takes place between both axes (Fig. 1a) and thus is affected by ε in both lattice directions. In Fig. S3, we present a complete orbital resolved analysis of exchange parameters as a function of strain. Application of strain in *a* can be used to precisely modify the influence of the t$_{2g}$-t$_{2g}$ mechanism (Fig. 1f), which is reduced by minimizing the overlap between the in-plane *d$_{xy}$, p$_x$ and p$_y$* orbitals. Thus, an expansion of the *a* lattice parameter results in a decrease of the AFM contribution of J$_1$, favoring FM interactions. The opposite effect is encountered when applying strain along *b*, which modifies the t$_{2g}$-e$_g$ channels that include orbitals with *z* component. This mainly affects second and third neighbors, resulting in an enhancement of ferromagnetism when the system is compressed along *b*.

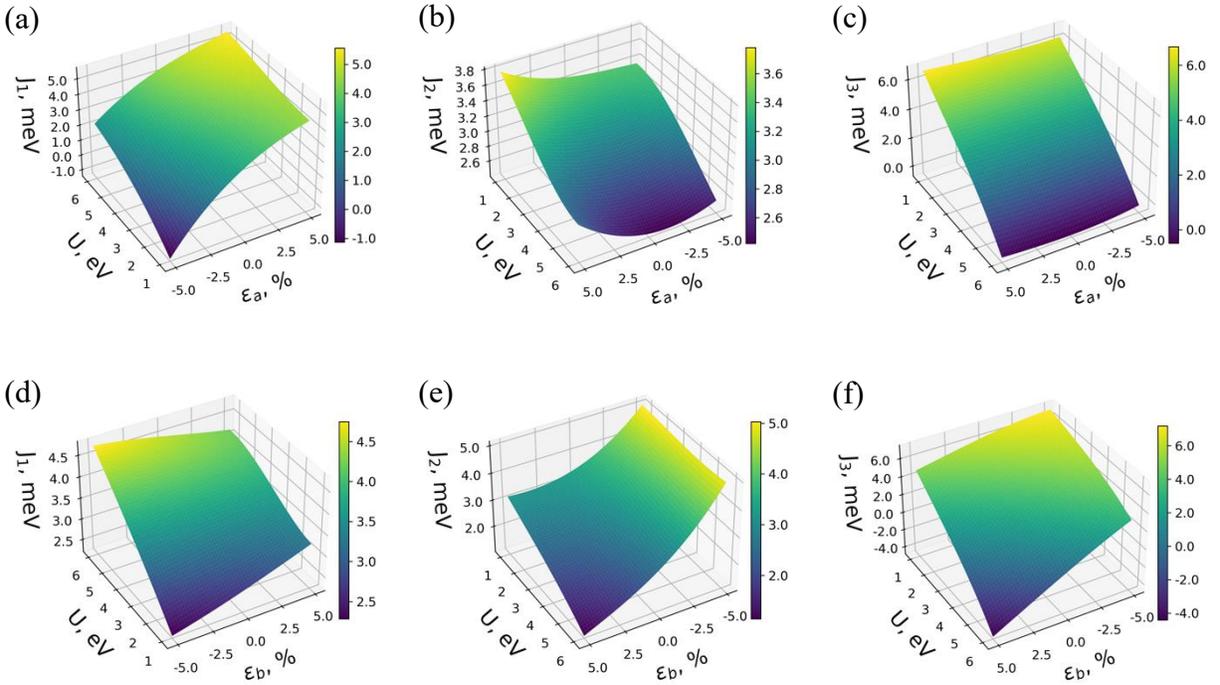

**Figure 2.** High density 3D surface plots of isotropic exchange parameters (J$_1$, J$_2$, J$_3$) as a function of strain (ε) and Hubbard U for CrSBr monolayer: (a-c) Uniaxial strain in *a* axis; (d-f) Uniaxial strain in *b* axis.

From magnetic exchange, the magnon dispersion relations are obtained via a Holfstein-Primakoff [45] transformation in the framework of linear spin-wave theory (LSWT), considering the bosonic operators terms up to second order. The resulting spin-wave Hamiltonian in reciprocal space is represented in eqs. 2 and 3:

$$\hat{H}_{SW} = E_0 + \sum_k \omega_k \cdot \hat{a}_k^\dagger \hat{a}_k \qquad (2)$$

$$\omega_k = 2S \sum_{n=1}^{3} \left( J_n \left(1 - \gamma_k^{(n)}\right) + J_n^z n_n \right) \qquad (3)$$

where $J_n$ ($J_n^z$) are isotropic (anisotropic) exchange parameters, $n_n$ the coordination numbers and $\gamma_k^{(n)}$ the structural factors of the different neighbors respectively.

Fig. 3 shows the effects of uniaxial strain on the magnon dispersion of CrSBr monolayer for U = 3 eV. As a result of the two magnetic Cr atoms in the unit cell, we can observe both acoustic and optical modes, which are degenerated along X-S and S-Y directions when a symmetric anisotropic Hamiltonian is considered. This degeneracy could be lifted by antisymmetric Dzyaloshinskii-Moriya interaction (DMI) induced by proximity effects, as DMI is negligible in this 2D ferromagnet.[42] By contrast, highly dispersive branches are present in the Γ-X and Γ-Y directions which are related to the two main reciprocal axes (Γ-X and Γ-Y correspond to *a* and *b* axis direction in real space, respectively). This leads to a selective control of the magnon dispersion with strain as can be observed in Fig 3a (uniaxial *a*) and 3b (uniaxial *b*), having a strong influence in the anisotropy gap at Γ,[46] which plays a crucial role to have long-range magnetic ordering at a finite temperature, thus preventing the Mermin-Wagner theorem.[47]

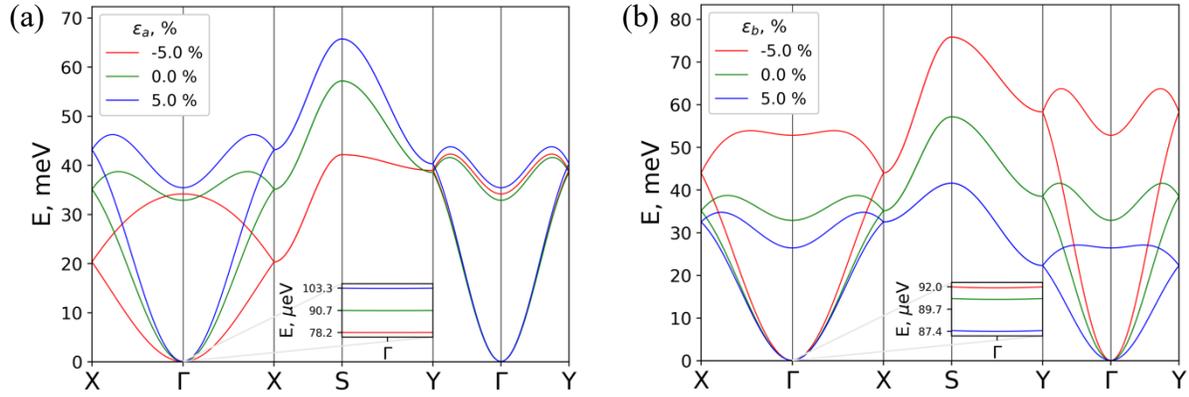

**Figure 3.** Magnon dispersion of CrSBr monolayer (U = 3 eV): (a) Uniaxial strain in *a* axis (b) Uniaxial strain in *b* axis. Inset: anisotropic gap at Γ point.

By applying renormalized spin-wave theory (RSWT), we calculate $T_C$. Fig. 4 presents the evolution of $T_C$ (a, c, d, f) and the gap at Γ point in spin wave spectrum (b, e) as a function of $\varepsilon$ and U. The calculated high-density 3D maps show a dramatic effect of uniaxial strain along *a* on the anisotropy gap, which increases (decreases) up to 14% (14%) with positive (negative) 5% strain for U = 3 eV (Fig. 3a and 4b). On the contrary, when strain is applied along *b*, the gap exhibits a more moderate decrease with positive strain, i.e. only by 4% under a 5% strain (Fig. 3b and 4e). This can be attributed to the unbalanced effect of both types of strain on the distorted octahedral coordination environment, which involves four ligands directly affected by strain along *a* (ac plane) versus two sulfur atoms along the *b* direction.

Regarding the dependence of the calculated $T_C$ with the anisotropy gap and the isotropic magnetic exchange interactions, one can find a clear competition between them when applying uniaxial deformation along the *a* axis (Fig. 4b, c). This is more evident at higher U values where $T_C$ drops dramatically despite the improvement of the anisotropy gap and $J_1$. The reason behind that is the growth of the population of the $e_g$ orbitals (Fig. S4) that activate the AFM exchange channel between the $d_{z2}$-$p_z$-$d_{z2}$ orbitals parallel to the *b* axis. Thus, $T_C$ follows the gap evolution for U < 4 eV, but at larger U $J_3$ becomes negative and starts to play a crucial role in destabilizing the FM configuration. On the other hand, our results show a good correlation between the gap and the critical temperature of CrSBr for compression/stretching applied in *b*. This is because in this case $J_3$ is directly controlled by the proposed experiment. The compression of the *b* parameter yields to a poorer effect of the $e_g$-$e_g$ AFM channel, thus making the FM configuration, mainly supported by $t_{2g}$-$e_g$ interactions, more stable. This results in a ~30% enhancement of $T_C$, predicting an upper limit for $T_C$ of 158 K for U =3 eV and ε = -5% along *b* (Fig. 4d). In the extreme region of U ~ 6 eV, for ε ~ 2%, and U ~ 5 eV, for ε ~ 3.5%, the system becomes antiferromagnetic along *b* and $T_C$ vanishes (blank region in Fig. 4f and Fig. S5).[48] This is also reflected in the magnon dispersion, where negative frequencies appear along Γ-Y path (See Fig S6). This behavior can be rationalized by the Goodenough-Kanamori rules[49] since superexchange pathway for $J_3$ becomes AFM when the angle Cr-S-Cr approaches 180º as long as the material is stretched along *b*.

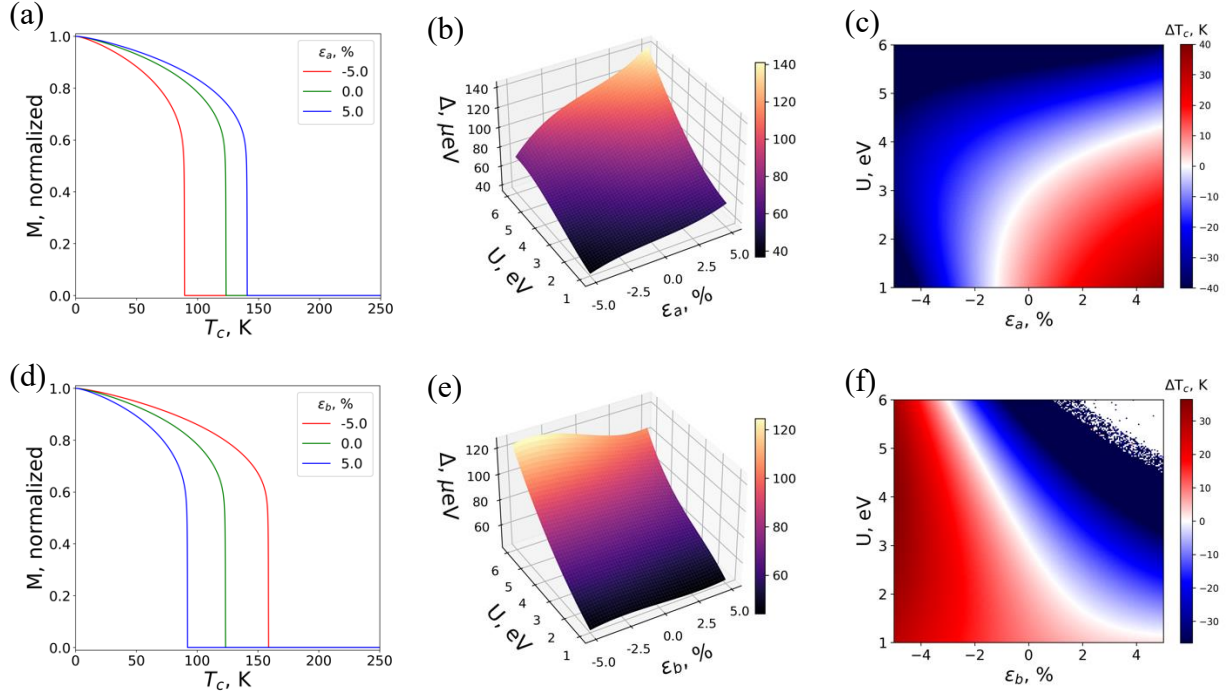

**Figure 4.** (a, d) Temperature dependence of Cr magnetic moment, (b, e) Gap at Γ point in magnon dispersion, (c, f) Curie temperature map; (a-b) Uniaxial strain in *a* axis, (d-f) Uniaxial strain in *b* axis. Relative point $T_c$ = 122 K corresponds to U = 3 eV and unstrained sample.

Finally, we evaluate the magnon dynamics by atomistic simulations based on Landau-Lifshitz-Gilbert (LLG) equation:[50, 51]

$$\frac{d\vec{m}}{dt} = -\gamma\mu_0 \vec{m} \times \vec{H} + \alpha \vec{m} \times \frac{d\vec{m}}{dt} \quad (4)$$

where $\vec{m}$ is the normalized magnetic moment of Cr atoms, $\vec{H}$ is the effective exchange field, $\mu_0$ is the permeability of vacuum, $\gamma$ is the gyromagnetic ratio and $\alpha$ is the Gilbert damping parameter that we set to a typical value of 0.01 for Cr-based 2D magnets.[52, 53] Our calculations start by perturbing the initial FM state with an oscillating magnetic field in a narrow region at the center of the sample for an ultrashort period of time (1 ps) with the objective of generating SWs.

Then, the SWs propagate as graphically shown by selected snapshots from our real-time real-space spin dynamics simulations (See Fig. S7).

The group velocity of SW propagation ($v$) is evaluated along the $a$ and $b$ crystallographic directions as a function of $\varepsilon$ and U (See Figs. 5 and S8). Note that the group velocity is determined for the strain region with FM order. Regarding its evolution, one can observe several abrupt changes (e.g. at $v_a = 3.2 \cdot 10^3$ and $4.2 \cdot 10^3$ m/s for $\varepsilon$ along $a$) that can be attributed to the deactivation of higher-frequency magnon modes when mechanically modifying the lattice coordinates. This effect can be corrected by choosing an optimal magnetic field frequency for each value of strain and U (Figs. S9 and S10). According to our dynamic simulations, environmental screening has an opposite effect on $v$ for $a$ and $b$ directions for both types of applied strain. In the case of $b$, the group velocity decreases with larger values of U. By contrast, higher Coulomb screening induces faster SWs along $a$. The effect of strain mainly affects $v$ in the same direction of the applied strain. Thus, for uniaxial strain along $a$, changes in $v_b$ are small –although they are more pronounced for larger values of U (See Fig. S8) – whereas $v_a$ can be significantly modified in the boundaries from -60% to 20% (See Fig. 5a).

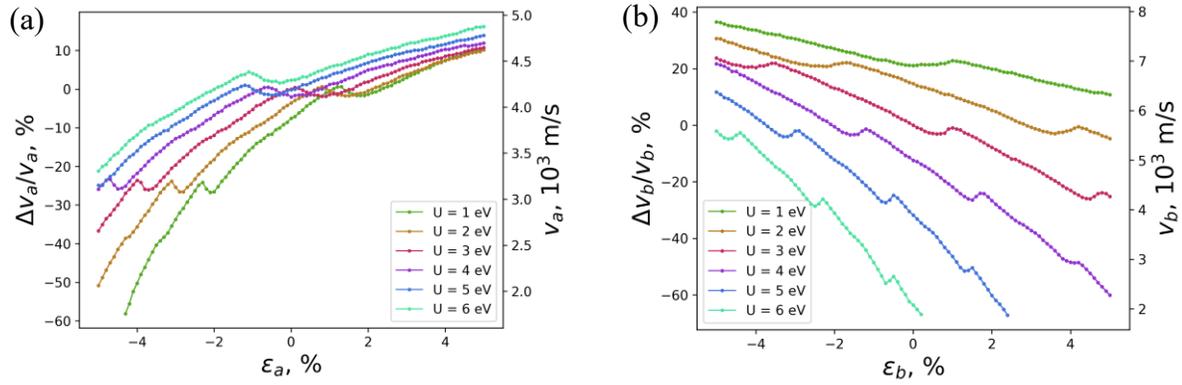

**Figure 5.** Group velocity in direction of *a* axis for uniaxial strain along *a* (a), and *b* axis for uniaxial strain along *b* (b). The reference points are chosen to be at ε = 0 %.

Our results show that strain-engineering tuning of group velocity is directly linked with changes of J values, which crucially depend on the orbital character of the possible exchange couplings. This is also evident when inspecting the slope of Γ-X and Γ-Y path in the magnon dispersion, which are parallel to *a* and *b*, respectively. Comparing Figs. 3 and 5, it is straightforward to realize that Γ-Y has a larger slope at 0% strain, which is translated into a faster $v_b = 5.8 \cdot 10^3$ m/s with respect to $v_a = 4.3 \cdot 10^3$ m/s according to the LLG dynamics simulations. Regarding the application of mechanical strain, we can see that the Γ-X slope is strongly tuned by uniaxial strain along *a*, which results in pronounced changes in $v_a$, whereas the Γ-Y path is controlled by uniaxial strain in *b*. This yields to faster SWs along the Cr-S-Cr 1D-chains. Overall, we can observe that along *a* (*b*), the SWs propagation is mostly influenced by $J_1$ ($J_3$), which can be microscopically controlled by uniaxial strain along those directions, with $J_2$ playing a minor role for both of them.

On the other hand, we have seen that the microscopic mechanisms that are responsible of the magnon dynamics can be tuned by varying the Coulomb interaction. In the 2D limit, dielectric properties are extremely sensitive to the environment, due to the weak screening that 2D materials can provide. In this sense, embedded layers, substrates, capping materials and encapsulation techniques perform an effective job quenching the electronic interactions.[54,55] In particular, substrates such as sapphire films or transition metal dichalchogenides, with a high dielectric constant are the most effective to decrease Coulombic interactions.[56] On the other hand, nanomaterials like hBN or graphene can provide a more efficient interface to reduce the U of the monolayer due to their flat nature, even offering the possibility of stacking more layers to produce a higher screening.[57,58] According to our predictions, this scenario will increase the group velocity, reaching the best performance for uniaxial compressive strain along *b*. In this sense, the more bulk is the capping material or substrate, the more screening is produced, though this effect usually saturates in the range of 3-4 layers.[59]

In summary, we have investigated the magnetic properties, magnon dispersion and spin dynamics of the air-stable 2D magnetic semiconductor CrSBr under uniaxial strain from first-principles. We provide a detailed microscopic understanding of the competing $t_{2g}$-$e_g$, $t_{2g}$-$t_{2g}$ and $e_g$-$e_g$ magnetic exchange channels that stabilize intralayer ferromagnetic coupling and shed light to the rational exploit of magnon straintronics in this intriguing quasi-1D material. Our calculations demonstrate that the magnon dynamics of CrSBr can be selectively tuned along the two main crystallographic directions (*a* and *b*) as a function of applied strain and environmental screening. Furthermore, we predict a strain-driven enhancement of $T_C$ ~30%, allowing the propagation of spin waves at higher temperatures.

AUTHOR INFORMATION

**Corresponding Author**

*E-mail: j.jaime.baldovi@uv.es. Phone +34-963444242.

**Author Contributions**

The manuscript was written through contributions of all authors. All authors have given approval to the final version of the manuscript. ‡These authors contributed equally.

**Notes**

The authors declare no competing financial interest.

ACKNOWLEDGMENT

The authors acknowledge the financial support from the European Union (ERC-2021-StG-101042680 2D-SMARTiES and FET-OPEN SINFONIA 964396), the Spanish MICINN (2D-HETEROS PID2020-117152RB-100, co-financed by FEDER, and Excellence Unit "María de Maeztu" CEX2019-000919-M) and the Generalitat Valenciana (grant CDEIGENT/2019/022, CIDEGENT/2018/004 and pre-doctoral fellowship GRISOLIAP/2021/038).

REFERENCES

(1) Demokritov, S. O.; Slavin, A. N. *Magnonics: From Fundamentals to Applications*; Springer Science & Business Media, 2012; Vol. 125.

(2) Lenk, B.; Ulrichs, H.; Garbs, F.; Münzenberg, M. The Building Blocks of Magnonics. *Physics Reports* **2011**, *507* (4–5), 107–136. https://doi.org/10.1016/j.physrep.2011.06.003.

(3) Kruglyak, V. v; Demokritov, S. O.; Grundler, D. Magnonics. *Journal of Physics D: Applied Physics* **2010**, *43* (26), 264001. https://doi.org/10.1088/0022-3727/43/26/264001.

(4) Nikitov, S. A.; Kalyabin, D. v; Lisenkov, I. v; Slavin, A.; Barabanenkov, Y. N.; Osokin, S. A.; Sadovnikov, A. v; Beginin, E. N.; Morozova, M. A.; Filimonov, Y. A.; Khivintsev, Y. v; Vysotsky, S. L.; Sakharov, V. K.; Pavlov, E. S. Magnonics: A New Research Area in Spintronics and Spin Wave Electronics. *Physics-Uspekhi* **2015**, *58* (10), 1002–1028. https://doi.org/10.3367/UFNe.0185.201510m.1099.

(5) Chumak, A. v.; Vasyuchka, V. I.; Serga, A. A.; Hillebrands, B. Magnon Spintronics. *Nature Physics* **2015**, *11* (6), 453–461. https://doi.org/10.1038/nphys3347.

(6) Rodin, A.; Trushin, M.; Carvalho, A.; Castro Neto, A. H. Collective Excitations in 2D Materials. *Nature Reviews Physics* **2020**, *2* (10), 524–537. https://doi.org/10.1038/s42254-020-0214-4.

(7) Neusser, S.; Grundler, D. Magnonics: Spin Waves on the Nanoscale. *Advanced Materials* **2009**, *21* (28), 2927–2932. https://doi.org/10.1002/adma.200900809.

(8) Huang, B.; Clark, G.; Navarro-Moratalla, E.; Klein, D. R.; Cheng, R.; Seyler, K. L.; Zhong, D.; Schmidgall, E.; McGuire, M. A.; Cobden, D. H.; Yao, W.; Xiao, D.; Jarillo-Herrero, P.; Xu, X. Layer-Dependent Ferromagnetism in a van Der Waals Crystal down to the Monolayer Limit. *Nature* **2017**, *546* (7657), 270–273. https://doi.org/10.1038/nature22391.

(9) Lee, J.-U.; Lee, S.; Ryoo, J. H.; Kang, S.; Kim, T. Y.; Kim, P.; Park, C.-H.; Park, J.-G.; Cheong, H. Ising-Type Magnetic Ordering in Atomically Thin $FePS_3$. *Nano Letters* **2016**, *16* (12), 7433–7438. https://doi.org/10.1021/acs.nanolett.6b03052.

(10) Jiang, S.; Li, L.; Wang, Z.; Mak, K. F.; Shan, J. Controlling Magnetism in 2D CrI3 by Electrostatic Doping. *Nature Nanotechnology* **2018**, *13* (7), 549–553. https://doi.org/10.1038/s41565-018-0135-x.

(11) Suárez Morell, E.; León, A.; Miwa, R. H.; Vargas, P. Control of Magnetism in Bilayer $CrI_3$ by an External Electric Field. *2D Materials* **2019**, *6* (2), 025020. https://doi.org/10.1088/2053-1583/ab04fb.

(12) Cenker, J.; Sivakumar, S.; Xie, K.; Miller, A.; Thijssen, P.; Liu, Z.; Dismukes, A.; Fonseca, J.; Anderson, E.; Zhu, X.; Roy, X.; Xiao, D.; Chu, J.-H.; Cao, T.; Xu, X. Reversible Strain-Induced Magnetic Phase Transition in a van Der Waals Magnet. *Nature Nanotechnology* **2022**, *17* (3), 256–261. https://doi.org/10.1038/s41565-021-01052-6.

(13) Ramos, M.; Carrascoso, F.; Frisenda, R.; Gant, P.; Mañas-Valero, S.; Esteras, D. L.; Baldoví, J. J.; Coronado, E.; Castellanos-Gomez, A.; Calvo, M. R. Ultra-Broad Spectral Photo-Response in FePS3 Air-Stable Devices. *npj 2D Materials and Applications* **2021**, *5* (1), 19. https://doi.org/10.1038/s41699-021-00199-z.

(14) Baltz, V.; Manchon, A.; Tsoi, M.; Moriyama, T.; Ono, T.; Tserkovnyak, Y. Antiferromagnetic Spintronics. *Reviews of Modern Physics* **2018**, *90* (1), 015005. https://doi.org/10.1103/RevModPhys.90.015005.

(15) He, Q. L.; Kou, X.; Grutter, A. J.; Yin, G.; Pan, L.; Che, X.; Liu, Y.; Nie, T.; Zhang, B.; Disseler, S. M.; Kirby, B. J.; Ratcliff II, W.; Shao, Q.; Murata, K.; Zhu, X.; Yu, G.; Fan, Y.; Montazeri, M.; Han, X.; Borchers, J. A.; Wang, K. L. Tailoring Exchange Couplings in Magnetic Topological-Insulator/Antiferromagnet Heterostructures. *Nature Materials* **2017**, *16* (1), 94–100. https://doi.org/10.1038/nmat4783.

(16) Sahalianov, I. Yu.; Radchenko, T. M.; Tatarenko, V. A.; Cuniberti, G.; Prylutskyy, Y. I. Straintronics in Graphene: Extra Large Electronic Band Gap Induced by Tensile and Shear Strains. *Journal of Applied Physics* **2019**, *126* (5), 054302. https://doi.org/10.1063/1.5095600.

(17) Miao, F.; Liang, S.-J.; Cheng, B. Straintronics with van Der Waals Materials. *npj Quantum Materials* **2021**, *6* (1), 59. https://doi.org/10.1038/s41535-021-00360-3.

(18) Roldán, R.; Castellanos-Gomez, A.; Cappelluti, E.; Guinea, F. Strain Engineering in Semiconducting Two-Dimensional Crystals. *Journal of Physics: Condensed Matter* **2015**, *27* (31), 313201. https://doi.org/10.1088/0953-8984/27/31/313201.

(19) Castellanos-Gomez, A.; Roldán, R.; Cappelluti, E.; Buscema, M.; Guinea, F.; van der Zant, H. S. J.; Steele, G. A. Local Strain Engineering in Atomically Thin $MoS_2$. *Nano Letters* **2013**, *13* (11), 5361–5366. https://doi.org/10.1021/nl402875m.

(20) Castellanos-Gomez, A.; Roldán, R.; Cappelluti, E.; Buscema, M.; Guinea, F.; van der Zant, H. S. J.; Steele, G. A. Local Strain Engineering in Atomically Thin $MoS_2$. *Nano Letters* **2013**, *13* (11), 5361–5366. https://doi.org/10.1021/nl402875m.

(21) Frisenda, R.; Drüppel, M.; Schmidt, R.; Michaelis de Vasconcellos, S.; Perez de Lara, D.; Bratschitsch, R.; Rohlfing, M.; Castellanos-Gomez, A. Biaxial Strain Tuning of the Optical Properties of Single-Layer Transition Metal Dichalcogenides. *npj 2D Materials and Applications* **2017**, *1* (1), 10. https://doi.org/10.1038/s41699-017-0013-7.

(22) Bukharaev, A. A.; Zvezdin, A. K.; Pyatakov, A. P.; Fetisov, Y. K. Straintronics: A New Trend in Micro- and Nanoelectronics and Material Science. *Uspekhi Fizicheskih Nauk* **2018**, *188* (12), 1288–1330. https://doi.org/10.3367/UFNr.2018.01.038279.

(23) Chen, X.; Qi, J.; Shi, D. Strain-Engineering of Magnetic Coupling in Two-Dimensional Magnetic Semiconductor CrSiTe3: Competition of Direct Exchange Interaction and Superexchange Interaction. *Physics Letters A* **2015**, *379* (1–2), 60–63. https://doi.org/10.1016/j.physleta.2014.10.042.

(24) Kou, L.; Tang, C.; Zhang, Y.; Heine, T.; Chen, C.; Frauenheim, T. Tuning Magnetism and Electronic Phase Transitions by Strain and Electric Field in Zigzag $MoS_2$ Nanoribbons. *The Journal of Physical Chemistry Letters* **2012**, *3* (20), 2934–2941. https://doi.org/10.1021/jz301339e.

(25) Wang, Y.; Wang, S.-S.; Lu, Y.; Jiang, J.; Yang, S. A. Strain-Induced Isostructural and Magnetic Phase Transitions in Monolayer $MoN_2$. *Nano Letters* **2016**, *16* (7), 4576–4582. https://doi.org/10.1021/acs.nanolett.6b01841.

(26) Sadovnikov, A. v.; Grachev, A. A.; Sheshukova, S. E.; Stognij, A. I.; Serokurova, A. I.; Nikitov, S. A. Magnon Straintronics for Tunable Spin-Wave Transport with YIG/GaAs and YIG/PZT Structures; **2020**; p 020105. https://doi.org/10.1063/5.0032060.

(27) Sadovnikov, A. v.; Grachev, A. A.; Serdobintsev, A. A.; Sheshukova, S. E.; Yankin, S. S.; Nikitov, S. A. Magnon Straintronics to Control Spin-Wave Computation: Strain

Reconfigurable Magnonic-Crystal Directional Coupler. *IEEE Magnetics Letters* **2019**, *10*, 1–5. https://doi.org/10.1109/LMAG.2019.2943117.

(28) Sadovnikov, A. V.; Grachev, A. A.; Sheshukova, S. E.; Sharaevskii, Yu. P.; Serdobintsev, A. A.; Mitin, D. M.; Nikitov, S. A. Magnon Straintronics: Reconfigurable Spin-Wave Routing in Strain-Controlled Bilateral Magnetic Stripes. *Physical Review Letters* **2018**, *120* (25), 257203. https://doi.org/10.1103/PhysRevLett.120.257203.

(29) Telford, E. J.; Dismukes, A. H.; Lee, K.; Cheng, M.; Wieteska, A.; Bartholomew, A. K.; Chen, Y.; Xu, X.; Pasupathy, A. N.; Zhu, X.; Dean, C. R.; Roy, X. Layered Antiferromagnetism Induces Large Negative Magnetoresistance in the van Der Waals Semiconductor CrSBr. *Advanced Materials* **2020**, *32* (37), 2003240. https://doi.org/10.1002/adma.202003240.

(30) Lee, K.; Dismukes, A. H.; Telford, E. J.; Wiscons, R. A.; Wang, J.; Xu, X.; Nuckolls, C.; Dean, C. R.; Roy, X.; Zhu, X. Magnetic Order and Symmetry in the 2D Semiconductor CrSBr. *Nano Letters* **2021**, *21* (8), 3511–3517. https://doi.org/10.1021/acs.nanolett.1c00219.

(31) López-Paz, S. A. et al. Dynamic Magnetic Crossover at the Origin of the Hidden-Order in van der Waals Antiferromagnet CrSBr. 1–12 (**2022**).

(32) Göser, O.; Paul, W.; Kahle, H. G. Magnetic Properties of CrSBr. *Journal of Magnetism and Magnetic Materials* **1990**, *92* (1), 129–136. https://doi.org/10.1016/0304-8853(90)90689-N.

(33) Wu, F.; Gutiérrez-Lezama, I.; López-Paz, S. A.; Gibertini, M.; Watanabe, K.; Taniguchi, T.; von Rohr, F. O.; Ubrig, N.; Morpurgo, A. F. Quasi-1D Electronic Transport in a 2D Magnetic Semiconductor. *Advanced Materials* **2022**, *34* (16), 2109759. https://doi.org/10.1002/adma.202109759.

(34) Yang, K.; Wang, G.; Liu, L.; Lu, D.; Wu, H. Triaxial Magnetic Anisotropy in the Two-Dimensional Ferromagnetic Semiconductor CrSBr. *Physical Review B* **2021**, *104* (14), 144416. https://doi.org/10.1103/PhysRevB.104.144416.

(35) Boix-Constant, C. et al. Probing the spin dimensionality in single-layer CrSBr van der Waals heterostructures by magneto-transport measurements (**2022**).

(36) Wang, H.; Qi, J.; Qian, X. Electrically Tunable High Curie Temperature Two-Dimensional Ferromagnetism in van Der Waals Layered Crystals. *Applied Physics Letters* **2020**, *117* (8), 083102. https://doi.org/10.1063/5.0014865.

(37) Telford, E. J.; Dismukes, A. H.; Lee, K.; Cheng, M.; Wieteska, A.; Bartholomew, A. K.; Chen, Y.; Xu, X.; Pasupathy, A. N.; Zhu, X.; Dean, C. R.; Roy, X. Layered Antiferromagnetism Induces Large Negative Magnetoresistance in the van Der Waals Semiconductor CrSBr. *Advanced Materials* **2020**, *32* (37), 2003240. https://doi.org/10.1002/adma.202003240.

(38) Rizzo, D. J.; McLeod, A. S.; Carnahan, C.; Telford, E. J.; Dismukes, A. H.; Wiscons, R. A.; Dong, Y.; Nuckolls, C.; Dean, C. R.; Pasupathy, A. N.; Roy, X.; Xiao, D.; Basov, D. N. Visualizing Atomically Layered Magnetism in CrSBr. *Advanced Materials* **2022**, 2201000. https://doi.org/10.1002/adma.202201000.

(39) Lee, K.; Dismukes, A. H.; Telford, E. J.; Wiscons, R. A.; Wang, J.; Xu, X.; Nuckolls, C.; Dean, C. R.; Roy, X.; Zhu, X. Magnetic Order and Symmetry in the 2D Semiconductor CrSBr. *Nano Letters* **2021**, *21* (8), 3511–3517. https://doi.org/10.1021/acs.nanolett.1c00219.

(40) Marzari, N.; Mostofi, A. A.; Yates, J. R.; Souza, I.; Vanderbilt, D. Maximally Localized Wannier Functions: Theory and Applications. *Reviews of Modern Physics* **2012**, *84* (4), 1419–1475. https://doi.org/10.1103/RevModPhys.84.1419.

(41) He, X.; Helbig, N.; Verstraete, M. J.; Bousquet, E. TB2J: A Python Package for Computing Magnetic Interaction Parameters. *Computer Physics Communications* **2021**, *264*, 107938. https://doi.org/10.1016/j.cpc.2021.107938.

(42) Scheie, A. et al. Spin waves and magnetic exchange Hamiltonian in CrSBr (**2022**).

(43) Guo, Y.; Zhang, Y.; Yuan, S.; Wang, B.; Wang, J. Chromium Sulfide Halide Monolayers: Intrinsic Ferromagnetic Semiconductors with Large Spin Polarization and High Carrier Mobility. *Nanoscale* **2018**, *10* (37), 18036–18042. https://doi.org/10.1039/C8NR06368K.

(44) Yang, K.; Wang, G.; Liu, L.; Lu, D.; Wu, H. Triaxial Magnetic Anisotropy in the Two-Dimensional Ferromagnetic Semiconductor CrSBr. *Physical Review B* **2021**, *104* (14), 144416. https://doi.org/10.1103/PhysRevB.104.144416.

(45) Holstein, T.; Primakoff, H. Field Dependence of the Intrinsic Domain Magnetization of a Ferromagnet. *Physical Review* **1940**, *58* (12), 1098–1113. https://doi.org/10.1103/PhysRev.58.1098.

(46) Lado, J. L.; Fernández-Rossier, J. On the Origin of Magnetic Anisotropy in Two Dimensional CrI$_3$. *2D Materials* **2017**, *4* (3), 035002. https://doi.org/10.1088/2053-1583/aa75ed.

(47) Mermin, N. D.; Wagner, H. Absence of Ferromagnetism or Antiferromagnetism in One- or Two-Dimensional Isotropic Heisenberg Models. *Physical Review Letters* **1966**, *17* (22), 1133–1136. https://doi.org/10.1103/PhysRevLett.17.1133.

(48) Cenker, J.; Sivakumar, S.; Xie, K.; Miller, A.; Thijssen, P.; Liu, Z.; Dismukes, A.; Fonseca, J.; Anderson, E.; Zhu, X.; Roy, X.; Xiao, D.; Chu, J.-H.; Cao, T.; Xu, X. Reversible Strain-Induced Magnetic Phase Transition in a van Der Waals Magnet. *Nature Nanotechnology* **2022**, *17* (3), 256–261. https://doi.org/10.1038/s41565-021-01052-6.

(49) Goodenough, J. B. *Magnetism and the Chemical Bond*; Hassell Street Press, **1963**; Vol. 1.

(50) LANDAU, L.; LIFSHITZ, E. On the Theory of the Dispersion of Magnetic Permeability in Ferromagnetic Bodies. In *Perspectives in Theoretical Physics*; Elsevier, **1992**; pp 51–65. https://doi.org/10.1016/B978-0-08-036364-6.50008-9.

(51) Gilbert, T. L. Classics in Magnetics A Phenomenological Theory of Damping in Ferromagnetic Materials. *IEEE Transactions on Magnetics* **2004**, *40* (6), 3443–3449. https://doi.org/10.1109/TMAG.2004.836740.

(52) Hiramatsu, R.; Miura, D.; Sakuma, A. First Principles Calculation for Gilbert Damping Constants in Ferromagnetic/Non-Magnetic Junctions. *AIP Advances* **2018**, *8* (5), 056016. https://doi.org/10.1063/1.5007255.

(53) Gilmore, K.; Idzerda, Y. U.; Stiles, M. D. Identification of the Dominant Precession-Damping Mechanism in Fe, Co, and Ni by First-Principles Calculations. *Physical Review Letters* **2007**, *99* (2), 027204. https://doi.org/10.1103/PhysRevLett.99.027204.

(54) Qiu, D. Y.; da Jornada, F. H.; Louie, S. G. Environmental Screening Effects in 2D Materials: Renormalization of the Bandgap, Electronic Structure, and Optical Spectra of Few-Layer Black Phosphorus. *Nano Letters* **2017**, *17* (8), 4706–4712. https://doi.org/10.1021/acs.nanolett.7b01365.

(55) Raja, A.; Chaves, A.; Yu, J.; Arefe, G.; Hill, H. M.; Rigosi, A. F.; Berkelbach, T. C.; Nagler, P.; Schüller, C.; Korn, T.; Nuckolls, C.; Hone, J.; Brus, L. E.; Heinz, T. F.; Reichman, D. R.; Chernikov, A. Coulomb Engineering of the Bandgap and Excitons in Two-Dimensional Materials. *Nature Communications* **2017**, *8* (1), 15251. https://doi.org/10.1038/ncomms15251.

(56) Qiu, D. Y.; da Jornada, F. H.; Louie, S. G. Environmental Screening Effects in 2D Materials: Renormalization of the Bandgap, Electronic Structure, and Optical Spectra of Few-Layer Black Phosphorus. *Nano Letters* **2017**, *17* (8), 4706–4712. https://doi.org/10.1021/acs.nanolett.7b01365.

(57) Trolle, M. L.; Pedersen, T. G.; Véniard, V. Model Dielectric Function for 2D Semiconductors Including Substrate Screening. *Scientific Reports* **2017**, *7* (1), 39844. https://doi.org/10.1038/srep39844.

(58) Peimyoo, N.; Wu, H.-Y.; Escolar, J.; de Sanctis, A.; Prando, G.; Vollmer, F.; Withers, F.; Riis-Jensen, A. C.; Craciun, M. F.; Thygesen, K. S.; Russo, S. Engineering Dielectric Screening for Potential-Well Arrays of Excitons in 2D Materials. *ACS Applied Materials & Interfaces* **2020**, *12* (49), 55134–55140. https://doi.org/10.1021/acsami.0c14696.

(59) Noori, K.; Cheng, N. L. Q.; Xuan, F.; Quek, S. Y. Dielectric Screening by 2D Substrates. *2D Materials* **2019**, *6* (3), 035036. https://doi.org/10.1088/2053-1583/ab1e06.